\documentclass[conference]{IEEEtran}
\IEEEoverridecommandlockouts

\usepackage{amsmath}
\usepackage{graphicx} 
\usepackage{epstopdf}
\usepackage{amssymb}
\usepackage{amsfonts}
\usepackage{amsthm}
\usepackage{cite}
\usepackage{bm,comment}
\usepackage{algorithm}
\usepackage{algorithmic}

\usepackage{subeqnarray}
\usepackage{subfigure}
\usepackage{multicol}
\usepackage{multirow}
\usepackage{diagbox}
\usepackage{slashbox}
\usepackage{stfloats}
\usepackage{float}
\usepackage{color}
\usepackage{cases}
\usepackage{CJK}
\usepackage{bm}

\usepackage[colorlinks,linkcolor=blue,citecolor=blue]{hyperref}
\usepackage{lipsum}

\usepackage{geometry}
\def\BibTeX{{\rm B\kern-.05em{\sc i\kern-.025em b}\kern-.08em
    T\kern-.1667em\lower.7ex\hbox{E}\kern-.125emX}}

\begin{document}
\title{Movable Antenna-Aided Hybrid Beamforming for Multi-User Communications}

\author{Yichi Zhang, Yuchen Zhang, Lipeng Zhu, Sa Xiao, Wanbin Tang, Yonina C. Eldar, \emph{Fellow, IEEE},\\ and Rui Zhang, \emph{Fellow, IEEE}

\thanks{

Yichi Zhang, Yuchen Zhang, Sa Xiao, and Wanbin Tang are with the National Key Laboratory of Wireless Communications, University of Electronic Science and Technology of China, Chengdu 611731, China (e-mail: \{yczhang,yc\_zhang\}@std.uestc.edu.cn, \{xiaosa,wbtang\}@uestc.edu.cn).

Lipeng Zhu is with the Department of Electrical and Computer Engineering, National University of Singapore,
Singapore 117583 (e-mail: zhulp@nus.edu.sg).

Yonina C. Eldar is with the Faculty of Mathematics and Computer Science, Weizmann Institute of Science, Rehovot 7610001, Israel (e-mail: yonina.eldar@weizmann.ac.il).

Rui Zhang is with The Chinese University of Hong Kong, Shenzhen, and Shenzhen Research Institute of Big Data, Shenzhen,
China 518172 (e-mail: rzhang@cuhk.edu.cn). He is also with the Department of Electrical and Computer Engineering, National
University of Singapore, Singapore 117583 (e-mail: elezhang@nus.edu.sg).
}

}
%

\maketitle

\begin{abstract}
In this correspondence, we propose a movable antenna (MA)-aided multi-user hybrid beamforming scheme with a sub-connected structure, where multiple movable sub-arrays can independently change their positions within different local regions. To maximize the system sum rate, we jointly optimize the digital beamformer, analog beamformer, and positions of
sub-arrays, under the constraints of
unit modulus, finite movable regions, and power budget. Due to the non-concave/non-convex objective function/constraints, as well as the highly coupled variables, the formulated problem is challenging to solve.
By employing fractional programming, we develop an alternating optimization framework to solve the problem via a combination of Lagrange multipliers, penalty method, and gradient descent. Numerical results
reveal that the proposed MA-aided hybrid beamforming scheme significantly improves the sum rate compared to its
fixed-position antenna (FPA) counterpart. Moreover, with sufficiently large movable regions, the proposed scheme with sub-connected MA arrays  even outperforms the fully-connected FPA array.
\end{abstract}

\section{Introduction}
Utilizing high-frequency millimeter-wave and/or terahertz bands for wireless communication has been recognized as an essential trend for advancing beyond 5G systems. This is attributed to the enhanced data rate available by exploiting their ultra-broad bandwidths. In order to compensate for the severe path loss in these bands, large antenna arrays are usually needed to provide high beamforming gains. Unfortunately, the heavy power consumption and high cost of radio frequency (RF) chains make fully-digital beamforming impractical. Hybrid beamforming \cite{Sub-connected-2,Hybrid_structure_2}, which leverages a small number of RF chains and an analog front end consisting of phase shifters (PSs), has been considered as a promising technique to achieve a good trade-off between hardware cost and communication performance. 

Based on the fabrication of PS enabled front end, the hybrid beamformer can be categorized
into fully-connected and sub-connected structures \cite{Sub-connected-3}. The fully-connected structure, with all RF chains connected to all antennas through PSs, allows more design degrees of freedom (DoFs). However, the excessive use of PSs entails a large cost. In comparison, by connecting each RF chain only with a portion of antennas, the sub-connected structure is more energy efficient and easier to implement  \cite{sub_connected_structure}. To improve the communication performance with a limited number of RF chains, different algorithms have been proposed \cite{Hybrid_structure_2,Sub-connected-2}. In \cite{Sub-connected-2}, the hybrid beamformer was optimized by approximating the fully-digital beamformer in an iterative manner. In \cite{Hybrid_structure_2}, the authors developed a general framework for hybrid beamforming design under various hardware architectures. However, both existing fully-connected and sub-connected hybrid beamforming schemes rely on fixed-position antennas (FPAs), which limit their communication performance because the spatial variation of wireless channels is not fully exploited at the transmitter/receiver.

{Recently, the movable antenna (MA), also known as fluid antenna system, has been proposed as a promising technology to enhance wireless communication performance \cite{MA_initial,FA_initial}. Specifically, an MA can flexibly tune its position and/or rotation to reconfigure the wireless channel towards a more favorable condition. Several prior works have investigated MA-aided systems for improving communication performance \cite{channel_model,FRV,MA_3,MA_4,MA_5}. The authors in \cite{channel_model} initially introduced a field-response based channel model for MA-aided systems. Based on this model, the work \cite{FRV} characterized the capacity of an MA-aided MIMO system. {{In \cite{MA_3,MA_4,MA_5}, the MA was generalized to multi-user systems. The authors in \cite{MA_4} extended MAs to secure communication by jointly optimizing the MA array geometry and beamforming vector.}} Most existing works considered that each MA is connected to an RF chain and a motor (for antenna movement) \cite{MA_initial,channel_model,FRV,MA_3,MA_4,MA_5}. However, implementing a fully-digital structure of an MA array with a large number of motors brings great challenges in system implementation.}

\begin{figure*}[!t]
\centering
\includegraphics[width=0.6\textwidth]{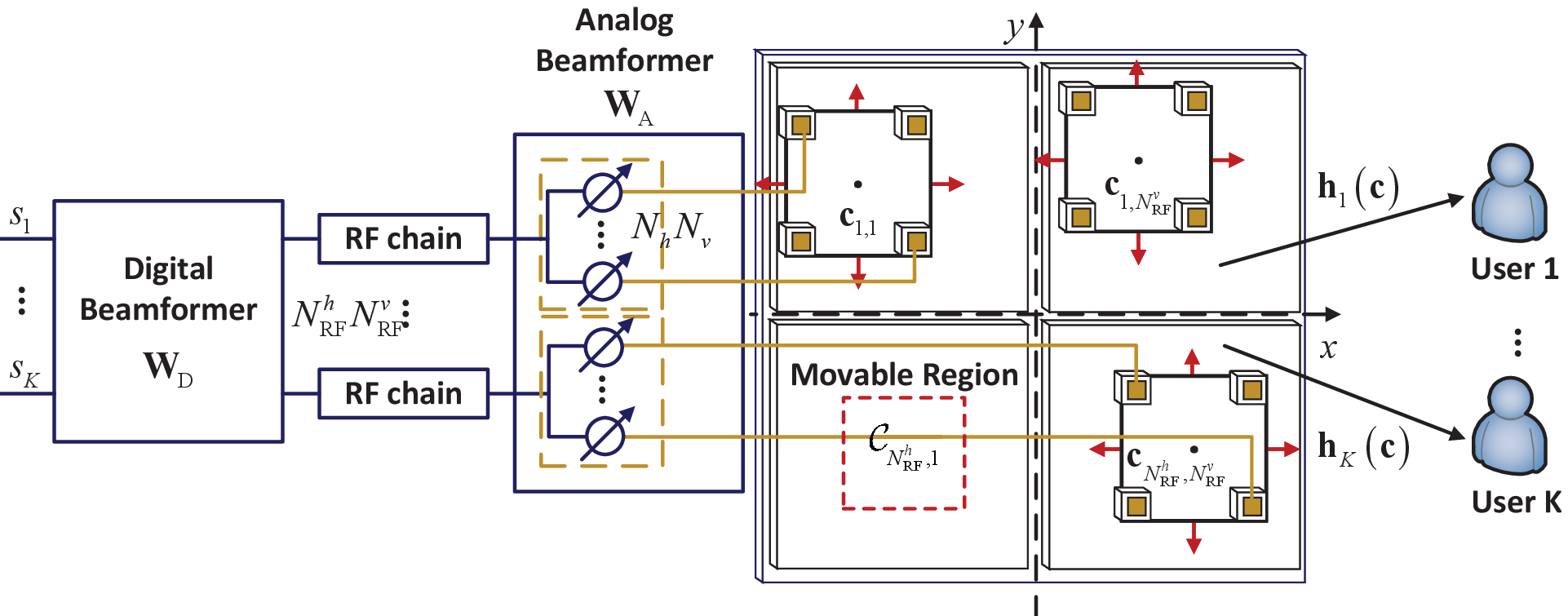}
\caption{MA-aided sub-connected hybrid beamforming structure.}
\label{system_model}
\vspace{-0.4cm}
\end{figure*}

In this correspondence, we propose a sub-connected structure of MA arrays. Instead of independently moving each antenna element, each sub-array is driven by a single motor such that the antennas therein can move collectively, which significantly reduces the implementation cost. Utilizing the proposed structure, we consider downlink multi-user communications with the aim of maximizing the system sum rate. The joint optimization of the hybrid beamforming and the positions of sub-arrays is a non-convex problem. Facilitated by fractional programming (FP), we propose an alternating optimization (AO) framework to solve the problem, by jointly applying Lagrange multipliers, penalty method, and gradient descent.
Numerical results demonstrate that our proposed MA-aided hybrid beamforming scheme outperforms its FPA counterpart. Moreover, under the condition of sufficiently large movable region size, the proposed scheme with sub-connected MA arrays yields a higher sum rate compared to the fully-connected FPA array.

\section{System Model}

We consider an MA-aided downlink multi-user multi-input single-output (MU-MISO) system as illustrated in Fig. \ref{system_model}. The system comprises a multi-antenna BS and $K$ single-antenna users indexed by $k\in\mathcal{K}\triangleq\{1,2,\cdots,K\}$. The BS adopts hybrid beamforming with a sub-connected structure \cite{Hybrid_structure_2,Sub-connected-2}, which consists of $N_{\rm{RF}}^h  N_{\rm{RF}}^v$ movable uniform planar arrays (UPAs), where $N_{\rm{RF}}^h$ and $N_{\rm{RF}}^v$ denote the number of UPAs along the horizontal and vertical directions, respectively. 

Each movable UPA consists of $N_hN_v$ antennas, where $N_h$ and $N_v$ represent the number of antennas along its horizontal and vertical directions, respectively. The total number of antennas is thus given by $N=N_{\rm{RF}}^h N_{\rm{RF}}^v N_h N_v$. The antennas within each UPA are first connected to a set of PSs individually and then to a dedicated RF chain, and they (together with their connected PSs) can move collectively. To avoid collision, each movable UPA is confined in a separate rectangular frame. For convenience, the UPA located in the $m$-th row and $n$-th column is indexed by $(m,n)$, $m=1,2,\cdots,N_{\rm{RF}}^h$ and $n=1,2,\cdots,N_{\rm{RF}}^v$. The index of the antenna in the $i$-th row and $j$-th column of a UPA is denoted as $(i,j)$, $i=1,2,\cdots,N_h$ and $j=1,2,\cdots,N_v$. The central position of the $(m,n)$-th UPA in the local coordinate system is denoted as ${\bf{c}}_{m,n}=(x_{m,n},y_{m,n})^T\in\mathcal{C}_{m,n}$, where $\mathcal{C}_{m,n}$ (enclosed by the red-dotted line in Fig. \ref{system_model}) represents the movable region. If the central point moves within this region, the UPA is guaranteed to move within the given frame. Additionally, for each UPA, we collect the relative positions of the antennas with respect to the central point in ${\bm{\delta}}=[{\bm{\delta}}_{1,1},{\bm{\delta}}_{1,2},\cdots,{\bm{\delta}}_{N_h,N_v}]\in \mathbb{R}^{2\times N_hN_v}$. The position of the $(i,j)$-th antenna within the $(m,n)$-th UPA is denoted by ${\bf{t}}_{m,n}^{i,j}={\bm{\delta}}_{i,j}+{\bf{c}}_{m,n}$. The location of the $k$-th user in its local coordinate system is represented by ${\bf{r}}_k$.

\subsection{Signal Model}

We consider quasi-static slow-fading channels. Let $L_k^t$ and $L_k^r$ represent the total number of transmit and receive channel paths from the BS to the $k$-th user, respectively. Define $\rho_{k,l}^{\chi}=[\sin\theta_{k,l}^{\chi}\cos\phi_{k,l}^{\chi},\cos\theta_{k,l}^{\chi}]$, where $\theta_{k,l}^{\chi}$ and $\phi_{k,l}^{\chi}$, $\chi\in\{t,r\}$, represent the elevation and azimuth angles of the $l$-th path, respectively. Let $\lambda$ denote the carrier wavelength. The transmit and receive field-response vectors (FRVs) between the $(i,j)$-th antenna within the $(m,n)$-th UPA and the $k$-th user are represented as \cite{channel_model,FRV}

\begin{equation}
{\bf{g}}_k\left({\bf{t}}_{m,n}^{i,j}\right)=\left[e^{j\frac{2\pi}{\lambda}\left({\bf{t}}_{m,n}^{i,j}\right)^T\rho_{k,l}^t}\right]^T_{1\leq l\leq L_k^t},
\end{equation}

\noindent and

\begin{equation}
{\bf{f}}_k=\left[e^{j\frac{2\pi}{\lambda}\left({\bf{r}}_{k}\right)^T\rho_{k,l}^r}\right]^T_{1\leq l\leq L_k^r},
\end{equation}

\noindent respectively. Define the path-response matrix (PRM) $\Sigma_k\in\mathbb{C}^{L_k^t\times L_k^r}$ as the response between all the transmit and receive channel paths. The channel vector between the $(m,n)$-th UPA and the $k$-th user is then given by

\begin{equation}
{\bf{h}}_{k}\left({\bf{t}}_{m,n}\right)={\bf{G}}_k\left({\bf{t}}_{m,n}\right)^H\Sigma_k{\bf{f}}_k,
\end{equation}

\noindent where ${\bf{G}}_k({\bf{t}}_{m,n})=[{\bf{g}}_k({\bf{t}}_{m,n}^{1,1}),{\bf{g}}_k({\bf{t}}_{m,n}^{1,2}),\cdots,{\bf{g}}_k({\bf{t}}_{m,n}^{N_h,N_v})]$ is the field-response matrix (FRM) at the $(m,n)$-th UPA. In this correspondence, we assume the above field-response channel model is known at the BS for each realization of multi-user channels. 

We collect the central positions of all movable UPAs in ${\bf{c}}=[{\bf{c}}_{1,1},{\bf{c}}_{1,2},\cdots,{\bf{c}}_{N_{\rm{RF}}^h,N_{\rm{RF}}^v}]$. The channel vector between the BS and the $k$-th user is given by

\begin{equation}\label{channel_vector}
 {\bf{h}}_{k}\left({\bf{c}}\right)=\left[{\bf{h}}_{k}\left({\bf{t}}_{1,1}\right)^T,{\bf{h}}_{k}\left({\bf{t}}_{1,2}\right)^T,\cdots,{\bf{h}}_{k}\left({\bf{t}}_{N_{{\rm{RF}}}^h,N_{{\rm{RF}}}^v}\right)^T\right]^T.
\end{equation}

\noindent Let ${\bf{s}}\in \mathbb{C}^K$ be the independent and identically distributed (i.i.d.) signal vector, with $\mathbb{E}(\bf{s}\bf{s}^H)={\bf{I}}_K$. The received signal at the $k$-th user is then given by

\begin{equation}\label{received_signal}
\begin{aligned}
y_k&={\bf{h}}_{k}\left({\bf{c}}\right)^H{\bf{W}}_{\rm{A}}{\bf{W}}_{\rm{D}}{\bf{s}}+n_k,\\
&=\underbrace{{\bf{h}}_{k}\left({\bf{c}}\right)^H{\bf{W}}_{\rm{A}}{{\bf{w}}_{k}s_{k}}}_{\rm{Desired ~signal}}+\underbrace{{\bf{h}}_{k}\left({\bf{c}}\right)^H{\bf{W}}_{\rm{A}}\sum_{k'\neq k}{{\bf{w}}_{k'}s_{k'}}}_{\rm{Interference}}+n_k,
\end{aligned}
\end{equation}

\noindent where ${\bf{W}}_{\rm{D}}=[{\bf{w}}_1,{\bf{w}}_2,\cdots,{\bf{w}}_{K}]\in\mathbb{C}^{N_{\rm{RF}}^hN_{\rm{RF}}^v\times K}$ denotes the digital beamformer, ${\bf{W}}_{\rm{A}}\in\mathbb{C}^{N\times N_{\rm{RF}}^hN_{\rm{RF}}^v}$ denotes the analog beamformer,  and $n_k\sim\mathcal{CN}(0,\sigma_k^2)$ is zero-mean additive white Gaussian noise (AWGN) with noise power $\sigma_k^2$. Due to the sub-connected struture, we have\cite{Sub-connected-3}

\begin{equation}\label{unit_modulus_constraint}
    {\bf{W}}_{\rm{A}}={\rm{diag}}\left({\bf{p}}_{1,1},{\bf{p}}_{1,2},\cdots,{\bf{p}}_{N_{\rm{RF}}^h,N_{\rm{RF}}^v}\right),
\end{equation}

\noindent {where ${{p}}_{m,n}^{i,j}=e^{j\psi_{m,n}^{i,j}}$ is the $((i-1)N_h+j)$-th element of ${\bf{p}}_{m,n}$ realized by a unit-modulus PS. Here $\psi_{m,n}^{i,j}\in[0,2\pi)$ denotes the phase of the PS connected to the $(i,j)$-th antenna within the $(m,n)$-th UPA.} Based on $\eqref{received_signal}$, the achievable rate of the $k$-th user is

\begin{equation}\label{rate}
R_k=\log_2\left(1+\frac{\left|{\bf{h}}_{k}\left({\bf{c}}\right)^H{\bf{W}}_{\rm{A}}{\bf{w}}_k\right|^2}{\sum_{k'\neq k}\left|{\bf{h}}_{k}\left({\bf{c}}\right)^H{\bf{W}}_{\rm{A}}{\bf{w}}_{k'}\right|^2+\sigma_k^2}\right).
\end{equation}

\subsection{Problem Formulation}
We aim to maximize the sum rate of all users jointly optimizing the central positions of UPAs ${\bf{c}}$, the analog beamformer ${\bf{W}}_{\rm{A}}$, and the digital beamformer ${\bf{W}}_{{\rm{D}}}$. The optimization problem is formulated as

\begin{subequations}\label{optimization_problem}
\begin{align}
\max_{{\bf{W}}_{\rm{D}},{\bf{W}}_{\rm{A}},{\bf{c}}}&\sum_{k=1}^KR_k\\
{\rm{s.t.}}~&{\bf{c}}_{m,n}\in\mathcal{C}_{m,n},~\forall m,n,\label{position_constraint}\\
&\left|\left|{\bf{W}}_{{\rm{A}}}{\bf{W}}_{{\rm{D}}}\right|\right|_F^2\leq P_{max},\label{power_budget}\\
&\left|p_{m,n}^{i,j}\right|=1,~\forall m,n,i,j\label{unit_modulus},
\end{align}
\end{subequations}
where $P_{max}$ denotes the power budget. Due to the non-concave/non-convex objective function/constraints, as well as the coupled variables, $\eqref{optimization_problem}$ is difficult to solve. Specifically, in contrast to the existing FPA-based hybrid beamforming designs \cite{Sub-connected-2,Hybrid_structure_2,hybrid_method}, the movable positions of UPAs introduce extra interdependency among variables, thus posing further challenges.
\section {Proposed Solution}
In this section, we develop a low-complexity algorithm to obtain a suboptimal solution for $\eqref{optimization_problem}$. Specifically, we first invoke the FP framework \cite{FP} to recast $\eqref{optimization_problem}$ into a more tractable form. Next, based on AO, we proceed to decompose the FP problem into three subproblems. Then, the penalty method is employed to address the unit modulus constraint. Finally, we propose a gradient decent method to optimize the positions of the UPAs.

\subsection{FP-Based Reformulation}
Let $\bm{\gamma}=[\gamma_1,\gamma_2,\cdots,\gamma_K]$ and $\bm{\omega}=[\omega_1,\omega_2,\cdots,\omega_K]$ be slack variables. To address the fractional forms in the objective function, we employ the FP framework in \cite{FP}, through which $\eqref{optimization_problem}$ is equivalently reformulated as

\begin{subequations}\label{FP}
\begin{align}
\max_{{\bf{W}}_{\rm{D}},{\bf{W}}_{\rm{A}},{\bf{c}},\bm{\gamma},\bm{\omega}}&\mathcal{L}=\sum_{k=1}^K\log\left(1+\gamma_k\right)-\gamma_k\notag\\
&\quad~~+\left(1+\gamma_k\right)\left(2\mathcal{R}\left\{\omega_k^*a_k\right\}-\left|\omega_k\right|^2b_k\right)\\
{\rm {s.t.}}~~&\gamma_k>0,\omega_k\in\mathbb{C},\forall k\in\mathcal{K},\\
&\eqref{position_constraint}, \eqref{power_budget}, \eqref{unit_modulus},
\end{align}
\end{subequations}
\noindent where, for fixed ${\bf{W}}_{\rm{D}}$, ${\bf{W}}_{\rm{A}}$, and ${\bf{c}}$, the optimal values of $\gamma_k$ and $\omega_k$ are given by

\begin{equation}\label{gamma}
\gamma_k^*=\frac{\left|a_k\right|^2}{\sum_{k'\neq k}\left|{\bf{h}}_{k}\left({\bf{c}}\right)^H{\bf{W}}_{\rm{A}}{\bf{w}}_{k'}\right|^2+\sigma_k^2}
\end{equation}
and 
\begin{equation}\label{omega}
\omega_k^*=\frac{a_k}{b_k},
\end{equation}
\noindent respectively, with $a_k={\bf{h}}_{k}\left({\bf{c}}\right)^H{\bf{W}}_{\rm{A}}{\bf{w}}_k$ and $b_k=\sigma_k^2+\sum_{k'=1}^K\left|{\bf{h}}_{k}\left({\bf{c}}\right)^H{\bf{W}}_{\rm{A}}{\bf{w}}_{k'}\right|^2$.

To decouple the variables $\{{\bf{W}}_{\rm{D}},{\bf{W}}_{\rm{A}},{\bf{c}},\bm{\gamma},\bm{\omega}\}$ in $\eqref{FP}$, we employ the AO framework to update each variable iteratively with other variables being fixed. Since the closed-form expressions of variables $\bm{\gamma}$ and $\bm{\omega}$ are provided in \eqref{gamma} and \eqref{omega}, respectively, we only need to update the other variables with given $\{\bm{\gamma},\bm{\omega}\}$ in each iteration. Define $\alpha_k=\log(1+\gamma_k)-\gamma_k-(1+\gamma_k)|\omega_k|^2\sigma_k^2$. Since $\alpha_k$ is only relevant to variables $\boldsymbol{\gamma}$ and $\boldsymbol{\omega}$, we omit it in the rest of this correspondence.

\subsection{Digital Beamformer Design}
Given $\{{\bf{W}}_{\rm{A}},{\bf{c}},\bm{\gamma},\bm{\omega}\}$, we seek the optimal value of the digital beamformer ${\bf{W}}_{\rm{D}}$. We recast the power constraint as

\begin{equation}
\left|\left|{\bf{W}}_{{\rm{A}}}{\bf{W}}_{{\rm{D}}}\right|\right|_F^2=N_vN_h\left|\left|{\bf{W}}_{{\rm{D}}}\right|\right|_F^2\leq P_{max}.
\end{equation}

\noindent Thus, ${\bf{W}}_{\rm{A}}$ and ${\bf{W}}_{\rm{D}}$ are decoupled in $\eqref{power_budget}$. The subproblem of digital beamformer design is then formulated as 

\vspace{-0.2cm}
\begin{subequations}\label{D_optimize}
\begin{align}
\max_{{\bf{W}}_{\rm{D}}}~&-\sum_{k=1}^K\mu_k\sum_{k'=1}^K\left|{\bm{\xi}}_{k}^H{\bf{w}}_{k'}\right|^2-2\mathcal{R}\left\{\bm{\beta}_k^H{\bf{w}}_k\right\}\\
{\rm {s.t.}}~&\left|\left|{\bf{W}}_{{\rm{D}}}\right|\right|_F^2\leq \frac{P_{max}}{N_vN_h},
\end{align}
\end{subequations}
\noindent where $\bm{\beta}_k^H=(1+\gamma_k)\omega_k^*{\bm{\xi}}_k^H$, $\bm{\xi}_k^H={\bf{h}}_k({\bf{c}})^H{\bf{W}}_{\rm{A}}$, and $\mu_k=(1+\gamma_k)|\omega_k^2|$. Let ${\bm{\Xi}}=\sum_{k=1}^K\mu_k\bm{\xi}_k\bm{\xi}_k^H$. According to the Lagrangian multipliers, the convex quadratic optimization problem admits the closed-form solution given by
\vspace{-0.2cm}
\begin{equation}\label{D_optimal}
{\bf{w}}_k^*\left(\lambda\right)=\left({\bm{\Xi}}+\lambda{\bf{I}}_{N_{\rm{RF}}^hN_{\rm{RF}}^v}\right)^{-1}\bm{\beta}_k,
\end{equation}
where $\lambda$ is a Lagrange multiplier such that $\lambda(||{\bf{W}}_{\rm{D}}||_F^2-\frac{P_{max}}{N_vN_h})=0$. That is, if $||{\bf{W}}_{\rm{D}}^*||_F^2=\sum_{k=1}^K||{\bf{w}}_k^*(0)||^2\leq\frac{P_{max}}{N_vN_h}$, $\lambda=0$. Otherwise, $\lambda>0$ can be found by a bisection search.

\subsection{Analog Beamformer Design}
With given $\{{\bf{W}}_{\rm{D}},{\bf{c}},\bm{\gamma},\bm{\omega}\}$, we set out to optimize ${\bf{W}}_{\rm{A}}$. Define $\tilde{{\bf{p}}}=[{\bf{p}}_{1,1}^T,{\bf{p}}_{1,2}^T\cdots,{\bf{p}}_{N_{\rm{RF}}^h,N_{\rm{RF}}^v}^T]^T$, $\tilde{{\bf{h}}}_{k,k'}=({\bf{w}}_{k'}\otimes {{\bf{I}}_{N_hN_v}}){\bf{h}}_{k}({\bf{c}})$, and $\tilde{{\bm{\beta}}}_k^H=(1+\gamma_k)\omega_k^*\tilde{{\bf{h}}}_{k,k}^H$. Based on the penalty method, the subproblem of analog beamformer design is formulated as
\vspace{-0.1cm}
\begin{subequations}\label{A_optimize}
\begin{align}
\max_{\tilde{\bf{p}},\bm{\phi}}~&\sum_{k=1}^K\left(-\mu_k\sum_{k'=1}^K\left|\tilde{{\bf{h}}}_{k,k'}^H\bm{\phi}\right|^2+2\mathcal{R}\left\{\tilde{{\bm{\beta}}}_k^H{\bm{\phi}}\right\}\right)\notag\\
&\quad~-\eta\left|\left|{\bm{\phi}}-\tilde{{\bf{p}}}\right|\right|^2_2\\
{\rm{s.t.}}&\left|\left[\tilde{\bf{p}}\right]_s\right|=1,\forall s=1,2,\cdots,N_{\rm{RF}}^hN_{\rm{RF}}^vN^hN^v,
\end{align}
\end{subequations}

\noindent where $\bm{\phi}$ is the introduced continuous variable and $\eta>0$ is the penalty parameter. Problem (14) can be solved via updating $\bf{\tilde{p}}$ and $\bm{\phi}$ iteratively. With given $\bf{\tilde{p}}$, the optimal solution of $\bm{\phi}$ is
\vspace{-0.1cm}
\begin{equation}\label{phi_optimal}
{\bm{\phi}}^*=\left[\tilde{{\bm{\Xi}}}\right]^{-1}\tilde{\bm{\beta}},
\end{equation}

\begin{algorithm}[t!]
    \renewcommand{\algorithmicrequire}{\bf{Input:}}
    \renewcommand{\algorithmicensure}{\bf{Output:}}
    \caption{Proposed algorithm for solving problem $\eqref{optimization_problem}$}
    \label{algorithm_1}
    \begin{algorithmic}[1]
    \REQUIRE${\bf{W}}_{\rm{D}}^{(0)},{\bf{W}}_{\rm{A}}^{(0)},{\bf{c}}^{(0)},\bm{\gamma}^{(0)},\bm{\omega}^{(0)},\left\{R_k^{(0)}\right\},\tilde{\kappa},\varepsilon,I_{max}$.
    \ENSURE ${{\bf{W}}_{\rm{D}}^{*},{\bf{W}}_{\rm{A}}^{*},{\bf{c}}^{*}}$.
    \STATE Set iteration index $t=1$.
    \REPEAT
    \STATE Update $\bm{\gamma}^{(t)}$ and $\bm{\beta}^{(t)}$ via \eqref{gamma} and \eqref{omega}, respectively.\
    \STATE Obtain ${\bf{W}}_{\rm{D}}^{(t)}$ via $\eqref{D_optimal}$.
    \STATE Obtain ${\bm{\phi}}^{(t)}$ via $\eqref{phi_optimal}$.
    \STATE Obtain ${\bf{W}}_{\rm{A}}^{(t)}$ via $\eqref{unit_modulus_constraint}$ and $\eqref{A_optimal}$.

    \FORALL{$m=1:N_{\rm{RF}}^h$}
    \FORALL{$n=1:N_{\rm{RF}}^v$}
    \STATE Calculate the gradient $\nabla_{{\bf{c}}_{m,n}}{\mathcal{L}({\bf{c}}_{m,n}^{(t)})}$ via $\eqref{gradient_value}$.
    \STATE Initialize the step size $\kappa=\tilde{\kappa}$.
    \REPEAT
    {\STATE Compute $\tilde{{\bf{c}}}_{m,n}={{\bf{c}}}_{m,n}^{(t)}+\kappa\nabla_{{\bf{c}}_{m,n}}{\mathcal{L}({\bf{c}}_{m,n}^{(t)})}$.
    \STATE Shrink the step size $\kappa\leftarrow\frac{\kappa}{2}$.
    \STATE Update ${{\bf{c}}}_{m,n}^{\left(t+1\right)}$ according to \eqref{Position_Update}.}
    \UNTIL{$\tilde{{\bf{c}}}_{m,n}\in\mathcal{C}_{m,n}$ and $\mathcal{L}(\tilde{{\bf{c}}}_{m,n})\geq\mathcal{L}({{\bf{c}}}_{m,n}^{(t)})$}.
    
    \ENDFOR
    \ENDFOR
    \STATE Calculate the sum rate $\sum_{k=1}^KR_k^{(t+1)}$ according to $\eqref{rate}$.
    \STATE Update $t=t+1$.
    \UNTIL{$\left|\sum_{k=1}^KR_k^{(t+1)}-R_k^{(t)}\right|<\varepsilon$} or the maximum iteration number $I_{max}$ is reached.
    \end{algorithmic}
\end{algorithm}

\noindent where $\tilde{\bm{\Xi}}=\sum_{k=1}^K\sum_{k'=1}^{K}({\mu}_k\tilde{{\bf{h}}}_{k,k'}\tilde{{\bf{h}}}_{k,k'}^H)+\eta{\bf{I}}_{N}$ and $\tilde{\bm{\beta}}=\sum_{k=1}^K\tilde{\bm{\beta}}_k+\eta\tilde{\bf{p}}$. Given $\bm{\phi}$, the optimization of $\bf{\tilde{p}}$ can be extracted from (14) as
\vspace{-0.1cm}
\begin{subequations}\label{analog_beamformer}
\begin{align}
\min_{\tilde{\bf{p}}}&\left|\left|{\bm{\phi}}-\tilde{{\bf{p}}}\right|\right|^2_2\\
{\rm{s.t.}}&\left|\left[\tilde{\bf{p}}\right]_s\right|=1,\forall s=1,2,\cdots,N_{\rm{RF}}^hN_{\rm{RF}}^vN^hN^v.
\end{align}
\end{subequations}

\noindent The optimal phase of $\tilde{\bf{p}}$ is given by
\vspace{-0.1cm}
\begin{equation}\label{A_optimal}
\arg\left\{\tilde{\bf{p}}\right\}=\arg\left\{{\bm{\phi}}\right\}.
\end{equation}

\subsection{MA Position Design}
Given $\{{\bf{W}}_{\rm{D}},{\bf{W}}_{\rm{A}},\bm{\gamma},\bm{\omega}\}$, the subproblem of MA position design can be decomposed into iteratively optimizing ${\bf{c}}_{m,n}$ with the other central points of UPAs fixed. The optimization problem is formulated as

\vspace{-0.4cm}
\begin{equation}\label{P_optimization}
\begin{aligned}
\max_{{\bf{c}}_{m,n}\in\mathcal{C}_{m,n}}&{\mathcal{L}({\bf{c}}_{m,n})}=\sum_{k=1}^K2\mathcal{R}\left\{{\bf{h}}_{k}\left({{{\bf{t}}_{m,n}}}\right)^H\bar{\bm{\beta}}_{k,m,n}\right\}\\
&-\mu_k\sum_{k'=1}^K\Bigg|\underbrace{\sum_{m=1}^{N_{\rm{RF}}^h}\sum_{n=1}^{N_{\rm{RF}}^v}{\bf{h}}_{k}\left({\bf{t}}_{m,n}\right)^H\bar{\bm{\xi}}_{k',m,n}}_{\bar{{h}}_{k,k'}\left({\bf{t}}_{m,n}\right)}\Bigg|^2,
\end{aligned}
\end{equation}
\noindent where $\bar{\bm{\beta}}_{k,m,n}=(1+\gamma_k)\omega_k^*\bar{\bm{\xi}}_{k,m,n}$, $\bar{\bm{\xi}}_{k,m,n}={{w}}_{k,m,n}{\bf{p}}_{m,n}$, with $w_{k,m,n}$ denoting the $((m-1)N_{\rm{RF}}^h+n)$-th element of ${\bf{w}}_{k}$.
Due to the complicated expression of ${\mathcal{L}({\bf{c}}_{m,n})}$, we propose a gradient descent method to optimize ${\bf{c}}_{m,n}$. In each iteration, the gradient can be calculated as

\vspace{-0.2cm}
\begin{equation}\label{gradient_value}
\begin{aligned}
\nabla_{{\bf{c}}_{m,n}}&{\mathcal{L}\left({\bf{c}}_{m,n}\right)}=2\sum_{k=1}^K\mathcal{R}\left\{\frac{\partial{{\bf{h}}_{k}\left({\bf{t}}_{m,n}\right)}}{\partial{{\bf{c}}_{m,n}}}\bar{\bm{\beta}}_{k,m,n}^*\right.\\
&\left.-\mu_k\sum_{k'=1}^K\frac{\partial{{\bf{h}}_{k}\left({\bf{t}}_{m,n}\right)}}{\partial{{\bf{c}}_{m,n}}}\bar{\bm{\xi}}_{k',m,n}^*{\bar{h}}_{k,k'}\left({\bf{t}}_{m,n}\right)\right\}.
\end{aligned}
\end{equation}
\noindent Then, the $(m,n)$-th central point in the $t$-th iteration is updated by moving along the gradient direction and checking whether the new point is still located within the feasible region, i.e.,

\vspace{-0.2cm}
\begin{equation}\label{Position_Update}
{\bf{c}}_{m,n}^{\left(t+1\right)}=\left\{
\begin{aligned}
\tilde{{\bf{c}}}_{m,n},&~{\rm{if}}~\tilde{{\bf{c}}}_{m,n}\in\mathcal{C}_{m,n} ~ {\rm{and}}~ \mathcal{L}(\tilde{{\bf{c}}}_{m,n})\geq\mathcal{L}({{\bf{c}}}_{m,n}^{(t)}),\\
{\bf{c}}_{m,n}^{\left(t\right)},&~{\rm{otherwise}},
\end{aligned}
\right.
\end{equation}

\noindent where $\tilde{{\bf{c}}}_{m,n}={\bf{c}}_{m,n}^{\left(t\right)}+\kappa\nabla_{{\bf{c}}_{m,n}}{\mathcal{L}({\bf{c}}_{m,n}^{(t)})}$ with $\kappa$ being the step size. Note that the performance of gradient descent heavily  depends on the step size. Thus, for each iteration, we initialize $\kappa$ with a large positive number and update $\kappa\leftarrow{\frac{\kappa}{2}}$ until finding a feasible point such that $\mathcal{L}(\tilde{{\bf{c}}}_{m,n})\geq\mathcal{L}({{\bf{c}}}_{m,n}^{(t)})$.

Algorithm $\ref{algorithm_1}$ summarizes the workflow for solving problem $\eqref{optimization_problem}$, which is guaranteed to converge as the objective function of the problem is non-decreasing over iterations and its optimal value is finite.


\section{Numerical Results}
In this section, numerical results are provided to evaluate the performance of our proposed scheme. Unless stated otherwise, we consider a scenario where a $16$-antenna BS operating at $f_c = 30$ GHz serves $K=4$ users. Besides, $N_{\rm{RF}}^h=N_{\rm{RF}}^v=N_h=N_v=2$. The relative position matrix is given by ${\bm{\delta}}=[\frac{-\lambda}{4},\frac{\lambda}{4},\frac{-\lambda}{4},\frac{\lambda}{4};\frac{\lambda}{4},\frac{\lambda}{4},\frac{-\lambda}{4},\frac{-\lambda}{4}]$, where $\lambda=0.01$ m represents the wavelength. The distance between the $k$-th user and the BS is uniformly distributed from 20 to 100 m, i.e., $d_k\sim\mathcal{U}[20,100]$. For each user, we assume an equal number of transmit and receive channel paths $L_k^t=L_k^r=L=6$. The PRM is defined as a diagonal matrix $\Sigma_k={\rm{diag}}\{\sigma_1,\sigma_2,\cdots,\sigma_L\}$, where $\sigma_{k,l}\sim\mathcal{CN}(0,(\rho_0d_k^{-\alpha})^2/L)$, $\rho_0=-40$ dB is the channel power gain at the reference distance 1 m, and $\alpha=2.8$ denotes the corresponding path loss exponent. The elevation and azimuth angles are assumed to follow the joint probability density function $f(\theta_{k,l}^{\chi},\phi_{k,l}^{\chi})=\frac{\cos\theta_{k,l}^{\chi}}{2\pi}$\cite{channel_model}, $\chi\in\{t,r\}$. For each UPA, the size of movable frame is $D\times D$. In addition, $P_{max}=10$ dBm, $\sigma_k^2=-80$ dBm, $D=2\lambda$, $\tilde{\kappa}=10$, $\varepsilon=10^{-3}$, and $I_{max}=200$.

To verify the effectiveness of the proposed scheme with Algorithm $\ref{algorithm_1}$, we consider two baselines for comparison: 1) Fixed array (FPA), sub-connected \cite{Sub-connected-3}. 2) Fixed array (FPA), fully-connected \cite{Sub-connected-3}. {Besides, by finding the optimal positions of all UPAs under the sub-connected structure within the whole region based on exhaustive search, the upper bound on the sum rate is also obtained for comparison.}

\begin{figure}[!t]
\vspace{-0.4cm}
\centering
\includegraphics[width=0.32\textwidth]{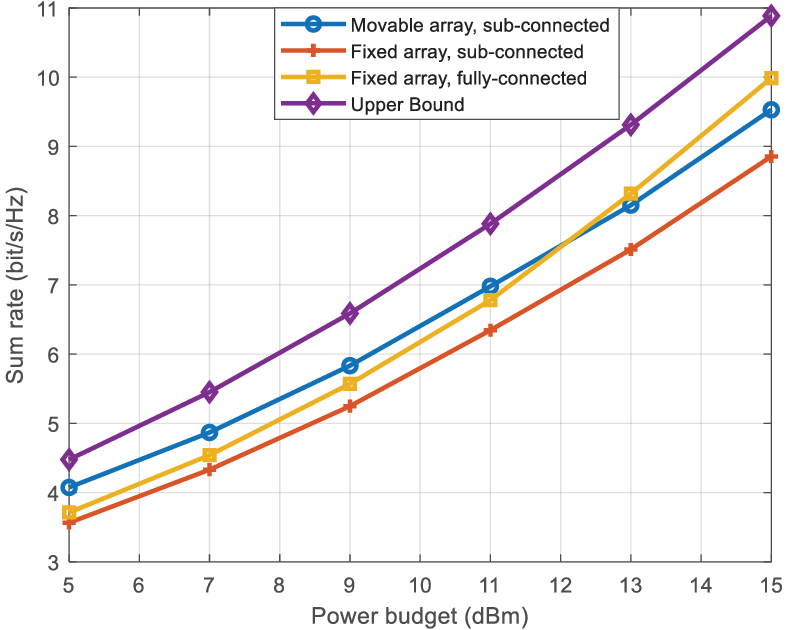}
\vspace{-0.3cm}
\caption{Sum-rate versus $P_{max}$ with $D=2\lambda$.}
\label{SNR_power}

\end{figure}
\begin{figure}[!t]
\vspace{-0.4cm}
\centering
\includegraphics[width=0.32\textwidth]{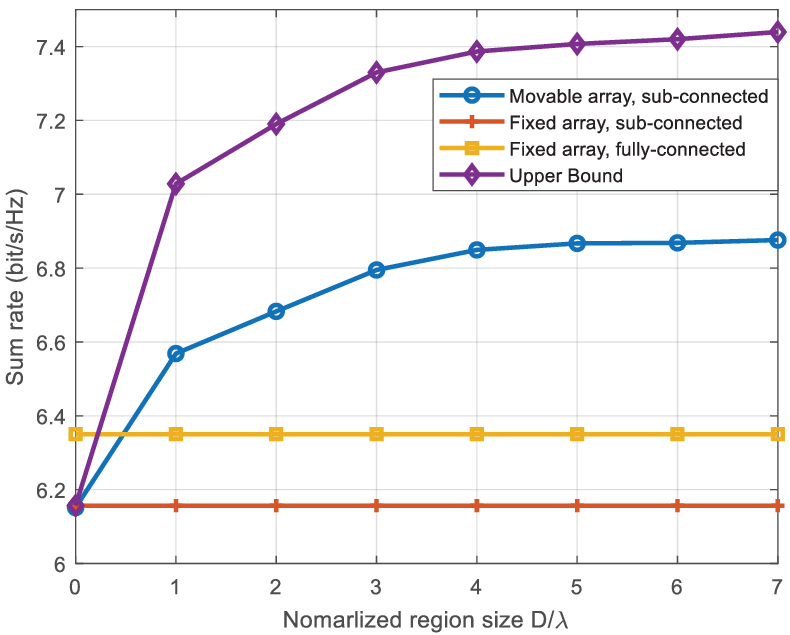}
\vspace{-0.3cm}
\caption{Sum-rate versus the normalized region size with $P_{max}=10$ dBm.}
\label{SNR_region}
\vspace{-0.4cm}
\end{figure}

Fig. $\ref{SNR_power}$ depicts the sum-rate versus the power budget for different schemes. It is observed that with the same power budget, our proposed algorithm achieves a higher sum-rate compared to its FPA counterpart. This suggests that the performance gain from MAs can compensate the performance loss of applying sub-connected structure. Moreover, when transmit power is sufficiently low, the MAs-aided sub-connected structure can even outperform the fully-connected FPAs, which indicates the efficacy of antenna moving capability.

Fig. $\ref{SNR_region}$ demonstrates the sum-rate versus $D/\lambda$. Note that the sum-rate increases as the normalized region size increases until reaching a constant value. The reason for it is that increasing region size provides more DoFs for the MAs to exploit the channel  spatial variation. For sufficiently large movable regions, the channels exhibit more evident correlation/periodicity in the spatial domain, and thus further increasing the region size has negligible improvement. Moreover, Fig. $\ref{SNR_region}$ also shows that with sufficiently large region size, our proposed scheme is superior to all baselines. {However, the performance gap from the upper bound is still substantial, due to that the gradient method is prone to converging to local optimum.}

\section{Conclusion}
This correspondence investigated MA-aided multi-user hybrid beamforming under the sub-connected structure. We studied the sum rate maximization problem by jointly designing the digital beamformer, analog beamformer, and movable UPAs' positions. To solve the non-convex optimization problem, we employed FP to transform it into a more tractable form and then developed an AO-based algorithm by applying the techniques of Lagrange multiplier, penalty method and gradient descent. Numerical results demonstrated the superiority of the proposed MA-aided sub-connected structure compared to the FPA-based system. Moreover, under certain conditions, the proposed scheme with sub-connected MA arrays even outperforms the fully-connected FPA array.

\end{document}